\documentclass[proceedings]{JHEP3}

\PrHEP{PrHEP hep2001}					
\conference{International Europhysics Conference on HEP}

\usepackage{epsfig}                   

\title{Low-Mass Lepton Pair Production in Pb-Au Collisions at 40 AGeV}

\author{\speaker{Sanja Damjanovic}	 			
	and Kirill Filimonov\thanks{Present address: LBNL, Berkeley,
USA } \hspace*{0.1cm} for the CERES/NA45 Collaboration\\ 	
        Univerity Heidelberg, Germany, Phylosophenweg 12, 69120\\					  	
        E-mail: \email{damjano@ceres1.physi.uni-heidelberg.de}}	

\abstract{
The CERES/NA45 experiment at the CERN SPS has previously measured 
$e^{+}e^{-}$ pair production in 160 AGeV Pb-Au collisions. In the mass 
region m$>$0.2 GeV/c$^{2}$, an enhancement of $2.7\pm0.4(statist.)\pm0.5(syst.)$
compared to the expectation from known hadronic decay sources was
observed. In the 40 AGeV data taken in 1999, an enhancement is again
found; a preliminary analysis gives the even larger value of $4.5\pm1.2(statist.)$.
The results are compared to theoretical model calculations based  on
$\pi^{+}\pi^{-}$ annihilation with a modified $\rho$-propagator.
}

\begin{document}

 \section{Introduction}
According to finite-temperature lattice QCD, strongly interacting
matter will at sufficiently high energy densities undergo a phase
transition from a state of hadronic constituents to quark matter, a
plasma of deconfined quarks and gluons. High-energy nucleus-nucleus
collisions have been used since about 15 years to experimentally
investigate this issue in the laboratory. Dileptons are a particularly 
attractive observable. In contrast to hadrons, they directly probe the 
early stages of the fireball evolution; the instantaneous emission
after creation and the absence of any final state interaction
conserves the primary information within the limits imposed by the
space-time folding over the emission period. In the low-mass region,
the thermal radiation is dominated by the decays of the light vector
mesons $\rho, \omega,$ and $\phi$. The $\rho$ is of particular
interest, due to its direct link to chiral symmetry and its short
lifetime of 1.3 fm/c; its in-medium behaviour around the critical
temperature T$_{c}$ for deconfinement should therefore reflect the
associated restoration of chiral symmetry. Experimentally, low-mass
dileptons are very much the domain of NA45/CERES, the only electron pair
spectrometer at the SPS. This article focuses on recent results from
CERES, obtained for Pb-Au collisions at 40 AGeV (1999 data).\\

\section{Experimental set-up}
The CERES spectrometer (Fig. 1) is optimized to measure low-mass electron pairs 
close to mid-rapidity (2.1$<$$\eta$$<$2.65) with full azimuthal coverage. 
Two silicon drift chambers (SIDC1,2), located 10 cm and 13.8 cm behind a 
segmented Au target, provide a precise angle measurement of the charged particles
and precise vertex reconstruction. Electron
\FIGURE{\epsfig{file=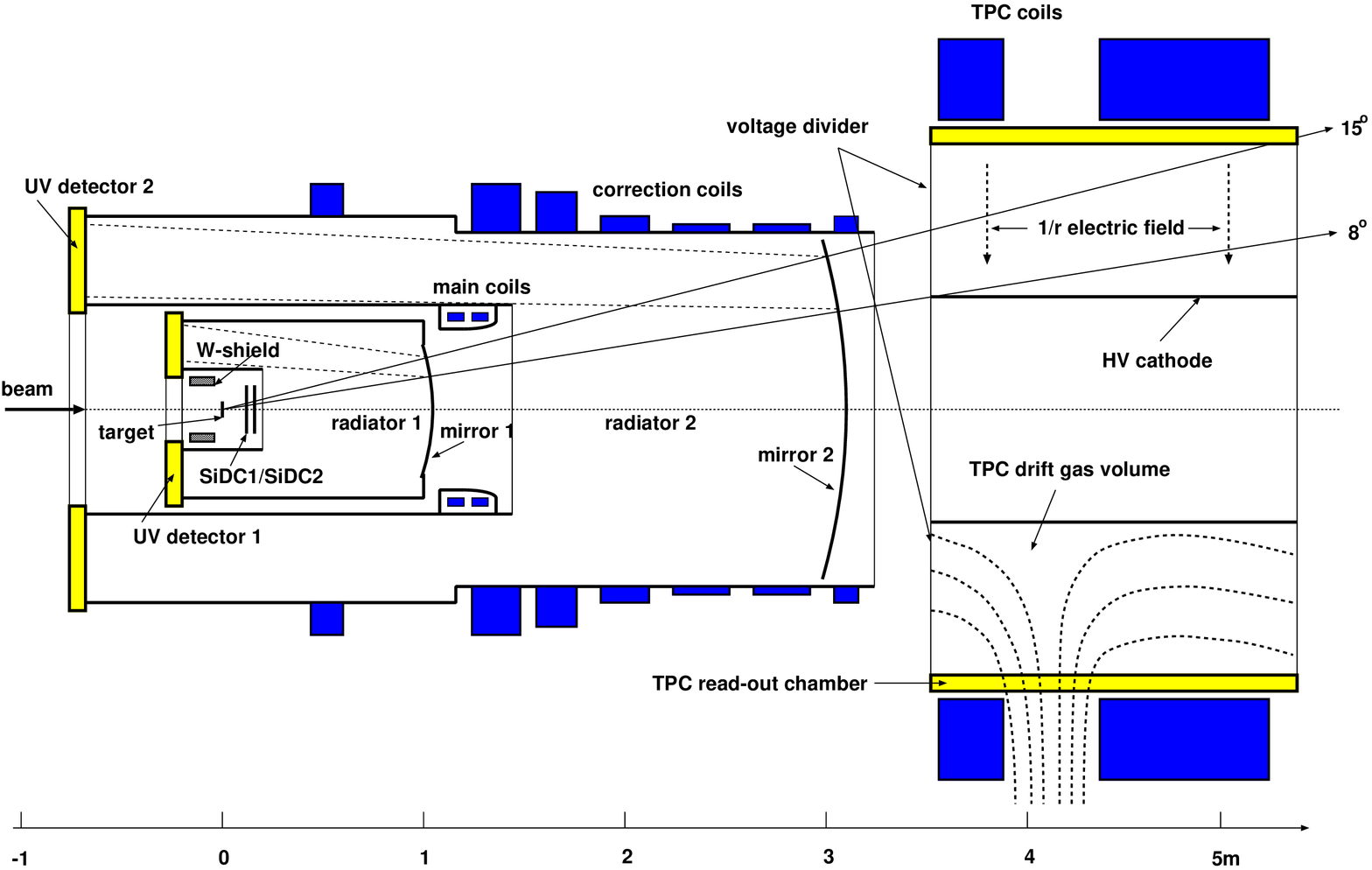,angle=270,width=0.89\textwidth}
\caption{\footnotesize The upgraded CERES spectrometer at the CERN SPS}
}
identification is done with two Ring Imaging Cherenkov detectors (RICH1,2)
operated at a high threshold $\gamma_{th}$=32 which rejects 95 $\%$ of all charged
hadrons. In the 1995/96 set-up[1], a suitably shaped magnetic field provided a 
well localized momentum kick between the two RICH detectors. A mass 
resolution of $\sim$5.5 $\%$ in the region of the $\rho/\omega$ was achieved. For the 
running period 1999/2000, the spectrometer was upgraded with a radial-drift 
Time Projection Chamber (TPC)[2] and a new magnetic field configuration. 
With an active length of 2 m, an outer diameter of 2.6 m, up to 20 space points 
for each track and a maximal radial field component of 0.5 T, the mass 
resolution is improved to a level of $\sim$2 $\%$ in the region of the 
$\rho/\omega$; only $\sim$4 $\%$ was, however, reached for the 1999 data set 
(the experiment was not quite ready yet). The two RICH detectors are
now used as an integral unit, resulting in improved electron efficiency (0.94
instead of 0.70 in 1995/1996) and improved rejection power.

\section{Analysis procedure}
The physics signal to be analyzed consists of electron-positron pairs
in the invariant-mass range 0.2$<$m$<$1.2 GeV/$c^2$. The abundance  of
these pairs is only of order 10$^{-5}$ relative to (i) hadrons and
(ii) photons. The double RICH system is powerful enough to essentially
control the problem (i) of hadron misidentification; in the 1999 data,
the remaining hadronic contamination in accidental matches between the
RICH's and the TPC is completely removed with the dE/dx information
from the TPC. The problem (ii) is much more severe. While the total
detector material up to the volume of RICH2 has been optimized down to
a level of $\sim$1 $\%$ of a radiation length, photons converted in the
target and in the SDC1 together with $\pi^0$-Dalitz decays still
exceed the number of high-mass pairs by a factor of $\sim$10$^{3}$.
Although the characteristics are different (small pair opening angles
and masses $<$0.2 GeV/$c^{2}$), limited track reconstruction efficiency and acceptance
leads to a {\it combinatorial high-mass background} for events in
which two or more low-mass pairs are only partially reconstructed.
In other words: low-mass pairs would have to be rejected on the level
of a factor of 10$^3$ to assure an (idealized) signal-to-background
ratio of 1/1. This is {\it the} problem of low-mass dilepton measurements.
\FIGURE{\epsfig{file=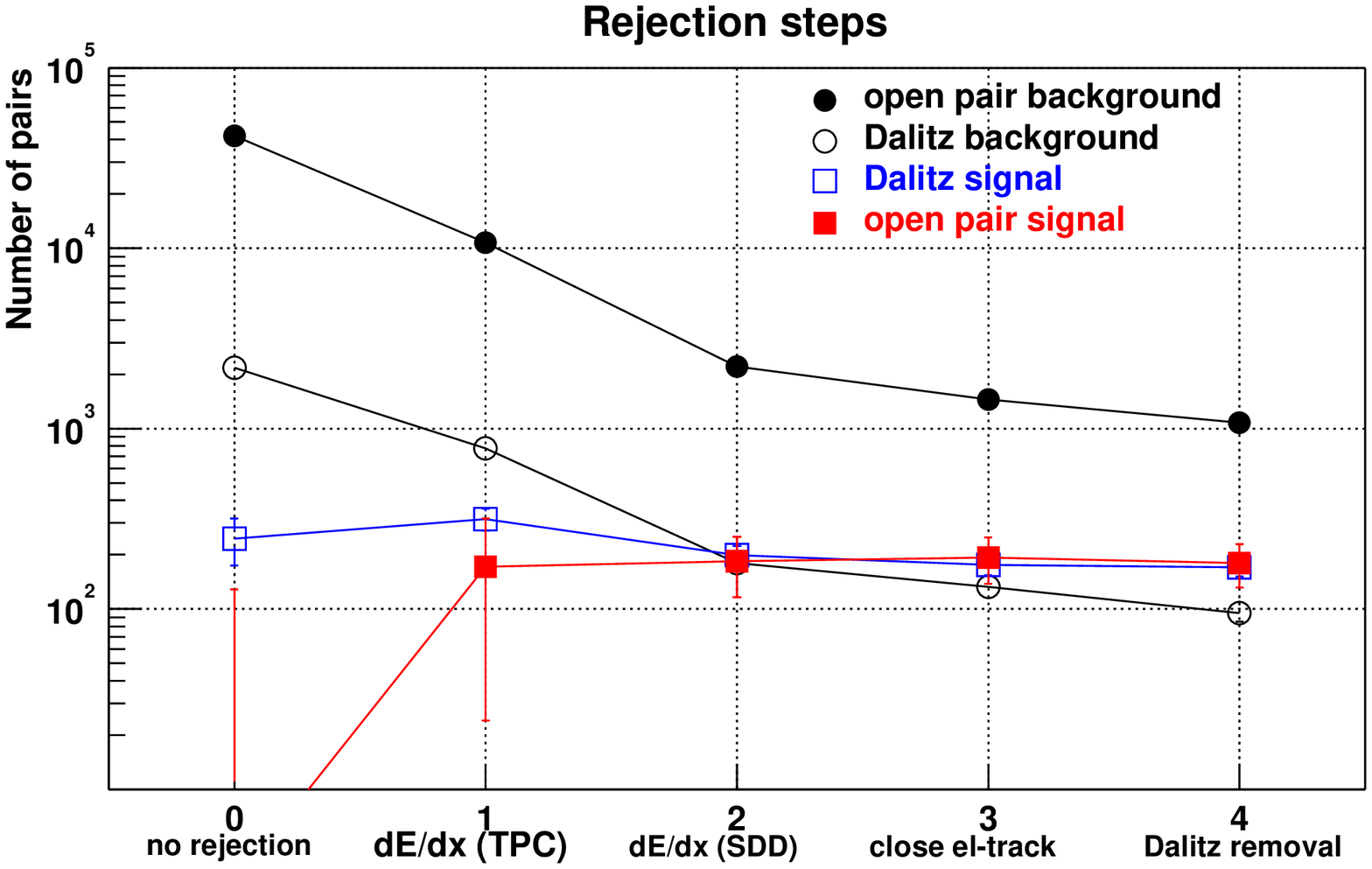, width=0.7\textwidth}
\caption{\footnotesize Evolution of the total number of pairs after various 
rejection steps, shown for the open-pair signal ($m_{ee}>0.2$ GeV/c$^2$), 
open-pair background ($m_{ee}>0.2$ GeV/c$^2$), Dalitz signal 
($m_{ee}<0.1$ GeV/c$^2$) and Dalitz background ($m_{ee}<0.1$
GeV/c$^2$) }
}
It is dealt with in the following way. A reduction of the combinatorial 
background by a factor of $>$10 is obtained by pairing only electron tracks 
with transverse momenta p$_{t}$$>$0.2 GeV/c. The most important remaining rejection
steps are illustrated in Fig. 2. Conversion and Dalitz pairs
with opening angles $<$10 mrad which are not recognized as two
individual electron rings in the RICH detectors are rejected 
by a double (correlated) dE/dx cut in the two SIDC's ( a factor of $\sim$5). 
Electron tracks which have a second SIDC-RICH electron candidate track within 
$<$70 mrad are also rejected. Finally, identified Dalitz pairs 
(m$<$0.2 GeV/c$^2$) are excluded from further combinatorics. Altogether, 
a rejection factor of $>$10$\cdot$10=100 is obtained. At the same time, the efficiency
losses for open pairs associated with the rejection cuts are less than a factor of 2, 
as illustrated by the positively identified Dalitz pairs with opening angles 
$>$35 mrad and masses $<$0.1 GeV/c$^2$ (Fig. 2). The physics signal,
i.e. high-mass pairs with m$>$0.2 GeV/c$^2$, is finally extracted
by subtracting the remaining like-sign pairs from the remaining
unlike-sign pairs as N$_{e^{+}e^{-}} - $2(N$_{e^{+}e^{+}}$$*$N$_{e^{-}e^{-}}$)$^{1/2}$.

\section{Results}
The measured $e^{+}e^{-}$ invariant-mass spectrum within the quoted rapidity acceptance, normalized to the number of
charged particles $<$N$_{ch}>$=$<$dN$_{ch}/$d$\eta>\cdot\Delta\eta$=115 in that acceptance, is shown in Fig. 3 [3]. The data
have been corrected for an overall pair reconstruction efficiency of $\epsilon$=0.05,
evaluated by overlaying open pairs of known characteristics on raw
events and simulating the whole set-up (``overlay Monte Carlo''). The
\FIGURE{\epsfig{file=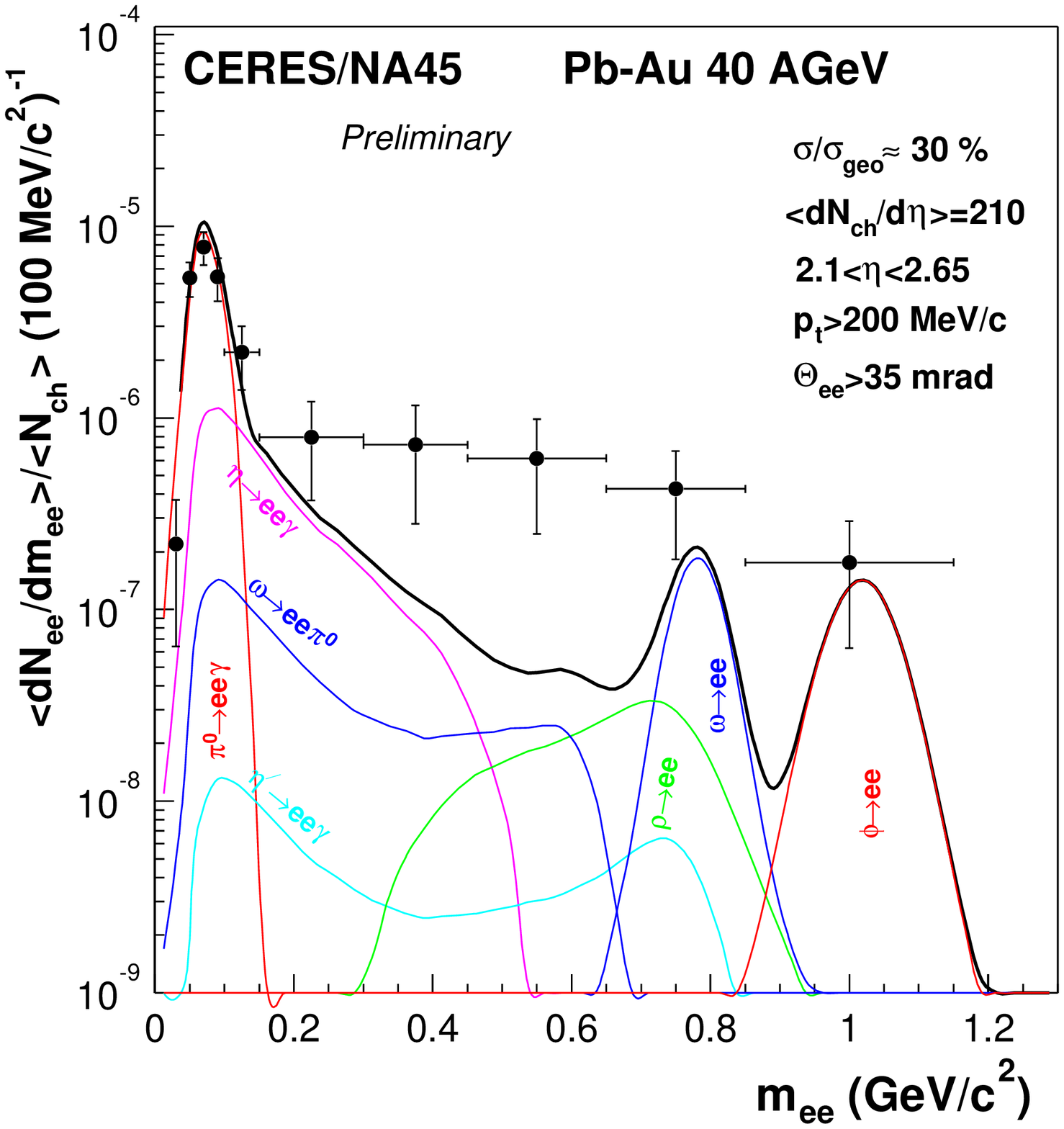, width=7.46cm,height=8.cm,clip=,bb=0 16 521 559}\hfill%
\epsfig{file=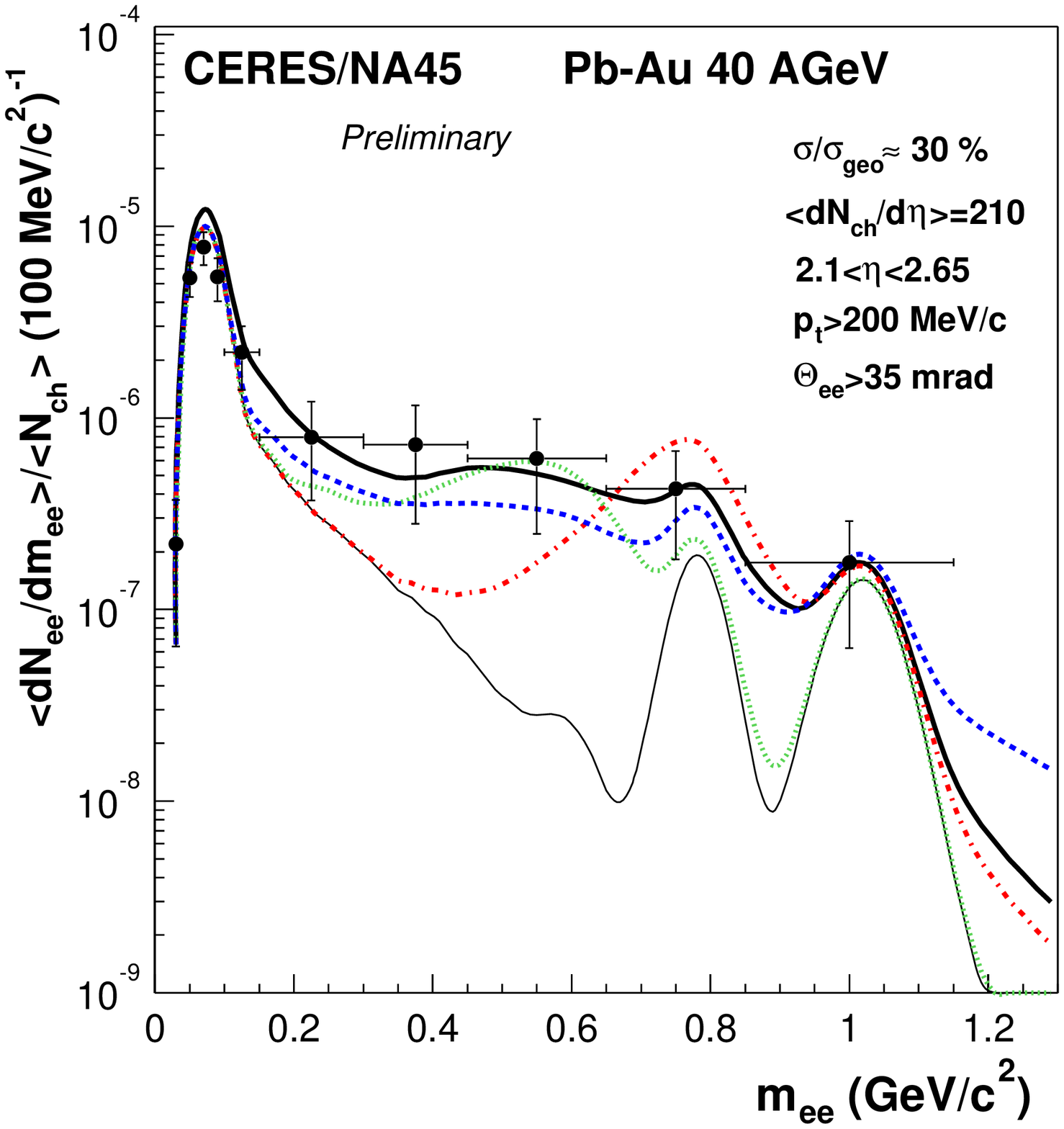, width=7.66cm,height=8.cm,clip=,bb=0 16 532 560
}\hfill%
\vspace*{-0.5cm}
\caption{Normalized invariant-mass spectra of
$e^{+}e^{-}$ pairs at 40 AGeV. Left panel: Comparison of the data to the
known hadronic decay sources, showing the individual contributions.
Right panel: Comparison of the data to model calculations [10], based either on
$\pi^{+}\pi^{-}$ annihilation with an unmodified $\rho$
(dashed-dotted), an in-medium dropping $\rho$ mass (dotted) and an
in-medium spreaded $\rho$ (thick solid), or on $q\overline{q}$
annihilation in the spirit of hadron-parton duality (dashed);
the sum of the hadronic decay sources, but without the $\rho$, is
shown separately (thin solid).}
}
unusually low number of 0.05 (instead of $>$0.2) is due to an only partially operational TPC in 
the start-up year 1999. A total number of about 8 Million events has
been analyzed. The number of net electron pairs entering into Fig. 3
is 249$\pm$28 with a signal-to-background ratio of 1/1 for m$<$0.2
GeV/c$^2$, and 180$\pm$48 with a signal-to-background
ratio of 1/6 for m$>$0.2 GeV/c$^2$. The stability of the two samples as a function of the
rejection steps, i.e. of varying widely the signal-to-background
ratio, can be recognized (within the error limits) in Fig. 2. The left
panel of Fig. 3 also contains a comparison to the known hadronic decay
sources, folded with a mass resolution of about 4 $\%$. The hadron
yields entering this ''decay cocktail'' include
the changes in Pb-Au collisions relative to pp. It should be stressed
that in the comparison study on pBe and pAu also done by CERES [4],
experimental data and the sum of the hadronic decay sources {\it agree
to within $<$20 $\%$}. Conversely here, a strong enhancement of the
data compared to the decay cocktail is seen, amounting for m$>$0.2
GeV/c$^2$ to a factor of $4.5\pm1.2(statist.)$. This is consistent with or
even larger than the enhancement of
$2.7\pm0.4(statist.)\pm0.5(syst.)$ reported by CERES for the full SPS
energy of 160 AGeV [5,6]. The excess appears to show
a threshold behaviour, i.e. to set in somewhere above 0.2 GeV/c$^2$
(roughly twice the $\pi$ rest mass). It is also found, for both
energies, to be most pronounced at low transverse momenta
p$_{t}^{ee}$$<$0.5 GeV/c.

More than 150 theoretical papers on the possible interpretation of the
CERES results have appeared in the last few years (for a review see
[7]). There seems to be a general consensus that one observes direct
radiation from the fireball, dominated by pion annihilation
$\pi^{+}\pi^{-}$$\rightarrow$$\rho$$\rightarrow$$e^{+}e^{-}$. The shape
of the $e^{+}e^{-}$ invariant-mass spectrum requires a strong medium
modification of the intermediate $\rho$. The two main contenders for
this are Brown-Rho scaling [8], reducing the mass (as a precursor of
chiral symmetry restoration), and a calculation of the $\rho$ spectral
density within the dense hadronic medium [7,9], spreading the
width. Both the time-averaged mass reduction and the spread of the
width are so large that the whole spectrum can be described, as a
parametrization, as if it were due to $q\overline{q}$ annihilation in
the spirit of hadron-parton duality. The quantitative description of
the data in this case requires an average temperature of 145 MeV at 40 
AGeV, compared to 170 AGeV ($\sim$T$_{c}$) at 160 AGeV [11].
As shown in the right panel of Fig. 3  the data quality clearly rules out an unmodified
$\rho$, i.e. a vacuum spectral function, while a more detailed
differentiation between the different medium scenarios is, within the
present errors, hardly possible. Much improved data have been taken in
2000 at 160 AGeV, and a further run in 2002 at a reduced SPS energy is
also under consideration.

\end{document}